\documentclass{article}
\usepackage{hiph-preprint}
\usepackage{exscale} 
\usepackage[latin1]{inputenc} 
\usepackage{amsmath,amssymb}
\usepackage{bbm,mathrsfs}
\usepackage{times}

\volnumber{22} \issuenumber{1} \edyear{2005}                             
\frompage{000} \topage{000}                                              
\recrevdate{November, 4, 2005}                                           
\def\rightmark{Leading-particle suppression and surface emission
in nucleus--nucleus collisions}
\def\leftmark{A.\ Dainese, C.\ Loizides and G.\ Pai\'c}

\newif\ifpdf
\ifx\pdfoutput\undefined
 \pdffalse
\else
 \pdftrue 
\fi

\ifpdf
\usepackage[pdftex]{graphicx}
\pdfoutput=1        
\pdfcatalog{/PageMode(/UseOutlines)} 
\else 
\usepackage[dvips]{graphicx}
\fi

\graphicspath{{./pics/}} 

\newif\ifcomment
\newif\ifbandw
\newcommand{\gsim}{\,{\buildrel > \over {_\sim}}\,}
\newcommand{\sqrtsNN}{\sqrt{s_{\scriptscriptstyle{{\rm NN}}}}}
\newcommand{\snn}{\sqrtsNN}
\newcommand{\av}[1]{\left\langle #1 \right\rangle}

\newcommand{\gev}{\mathrm{GeV}}

\newcommand{\fm}{\mathrm{fm}}

\newcommand{\AuAu}{\mbox{Au--Au}}

\newcommand{\dAu}{\mbox{d--Au}}

\newcommand{\RAA}{R_{\rm AA}}

\newcommand{\IAAa}{I^{\rm away}_{\rm AA}}

\newcommand{\pt}{p_{\rm t}}
\newcommand{\pta}{p^{\rm assoc}_{\rm t}}
\newcommand{\ptt}{p^{\rm trig}_{\rm t}}
\renewcommand{\d}{{\rm d}}

\newcommand{\Ref}[1]{Ref.~\cite{#1}}

\newcommand{\fig}[1]{fig.~\ref{#1}}

\newcommand{\colauthor}[1]{}

\title{Leading-particle suppression and surface emission
\mbox{in nucleus--nucleus collisions}} 
\authors{ 
{Andrea Dainese$^1$, Constantin Loizides$^{2}$ and Guy Pai\'c$^{3}$ %
\index{Dainese, A.} 
\index{Loizides, C.} 
\index{Pai\'c, G.} 
}\\[2.812mm]
{\normalsize
\hspace*{-8pt}$^1$ 
Universit\`a degli Studi di Padova and INFN, 
via Marzolo 8, 35131 Padova, Italy\\[0.2ex] 
\hspace*{-8pt}$^2$
Massachusetts Inst.~of Technology,
77 Mass Ave, Cambridge, MA, 02139, USA\\[0.2ex]
\hspace*{-8pt}$^3$
Instituto de Ciencias Nucleares, UNAM, Mexico City, Mexico 
}}

\abstract{
After a short summary of the predictions of the Parton Quenching Model 
(PQM) for the nuclear modification factor and its centrality dependence 
in \AuAu\ collisions at RHIC, we concentrate on back-to-back 
jet-like correlations at high transverse momentum. We illustrate 
how this probe is biased by the surface effect.    
}

\makeindex

\begin{document}
 
\maketitle

\section{Introduction}
\label{cl:intro}
The yield of high transverse-momentum ($\pt\gsim 5~\gev$) leading
particles in \AuAu\ collisions at the top RHIC energy, $\snn=200~\gev$,
is about a factor of five lower than expected from the measurements 
in pp collisions at the same energy~\cite{cl:starRAA,cl:phenixRAA}. 
Similarly, jet-like correlations on the azimuthally-opposite 
(`away') side of a high-$\pt$  trigger particle are suppressed 
by a factor of four to five, while the near-side correlation 
strength is almost unchanged~\cite{cl:starIAA}.
The absence of these effects in \dAu\ collisions at the same 
energy~\cite{cl:dAu} supports the partonic energy loss scenario: 
energetic partons, 
produced in initial hard scattering processes, lose energy as 
a consequence of the final-state interaction with the dense partonic 
matter created in \AuAu\ collisions. The dominant contribution to the 
energy loss is believed to originate from medium-induced gluon 
radiation (see Ref.~\cite{cl:eloss} and references therein). 

The Parton Quenching Model (PQM)~\cite{cl:PQM} combines 
the pQCD \mbox{BDMPS-Z-SW} framework for the probabilistic 
calculation of parton energy loss in extended partonic matter 
of given size and density~\cite{cl:carlosurs} with a realistic 
description of the collision overlap geometry in a static medium. 
We treat partons (and parton pairs) on a parton-by-parton basis 
using Monte Carlo techniques. Details on the quenching procedure 
and its application to high-$\pt$ data can be found in 
Ref.~\cite{cl:PQM}.
\enlargethispage{1cm}
The model has one single free parameter that sets the scale 
of the medium transport coefficient $\hat{q}$, the average 
transverse momentum squared transferred to the hard parton 
per unit path length, and, thus, the scale of the energy loss.
In this proceedings, we concentrate on the suppression phenomena, 
introduced above, and on the question to what extent the 
corresponding partonic probes penetrate the interior of the 
fireball.

\section{Suppression of leading particles and jet-like 
         two-particle correlations}
\label{cl:suppression}
\begin{figure}[b]
\begin{minipage}[t]{7.9cm} 
\includegraphics[width=8.0cm]{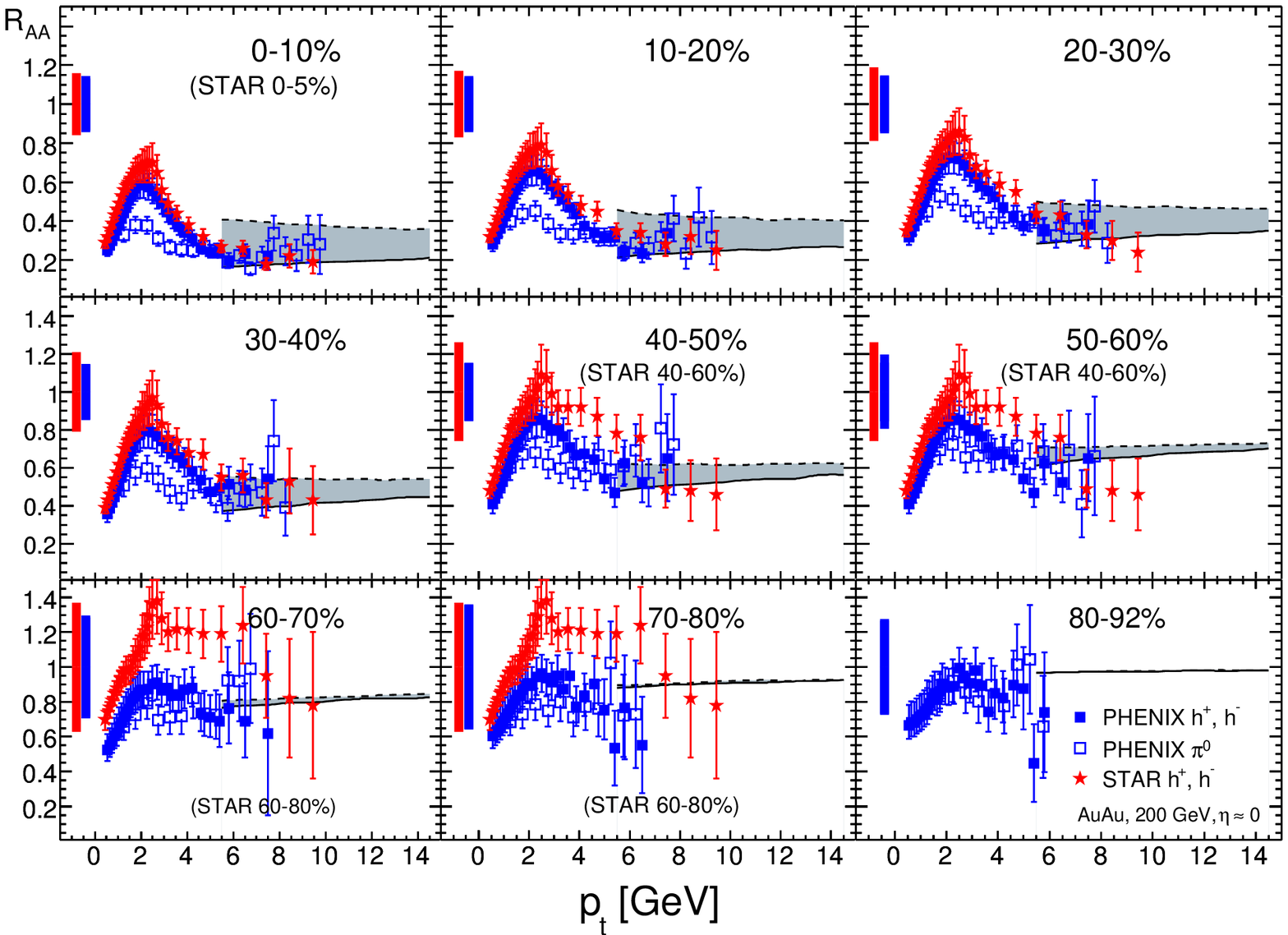}
\vspace{-0.9cm}
\caption[]{$\RAA(\pt)$ in \AuAu\ at $\snn=200~\gev$ for different centralities.
           Data are PHENIX charged hadrons and $\pi^0$~\cite{cl:phenixRAA}
           and STAR charged hadrons~\cite{cl:starRAA} with combined
           statistical and $\pt$-dependent systematic errors (bars on the 
           data points) and $\pt$-independent systematic errors (bars at 
           \mbox{$\RAA=1$}). The model band is the original PQM calculation 
           from~\Ref{cl:PQM}.
\label{cl:fig1}
} 
\end{minipage}
\hspace{0.2cm} 
\begin{minipage}[t]{5.2cm}
\includegraphics[width=4.8cm]{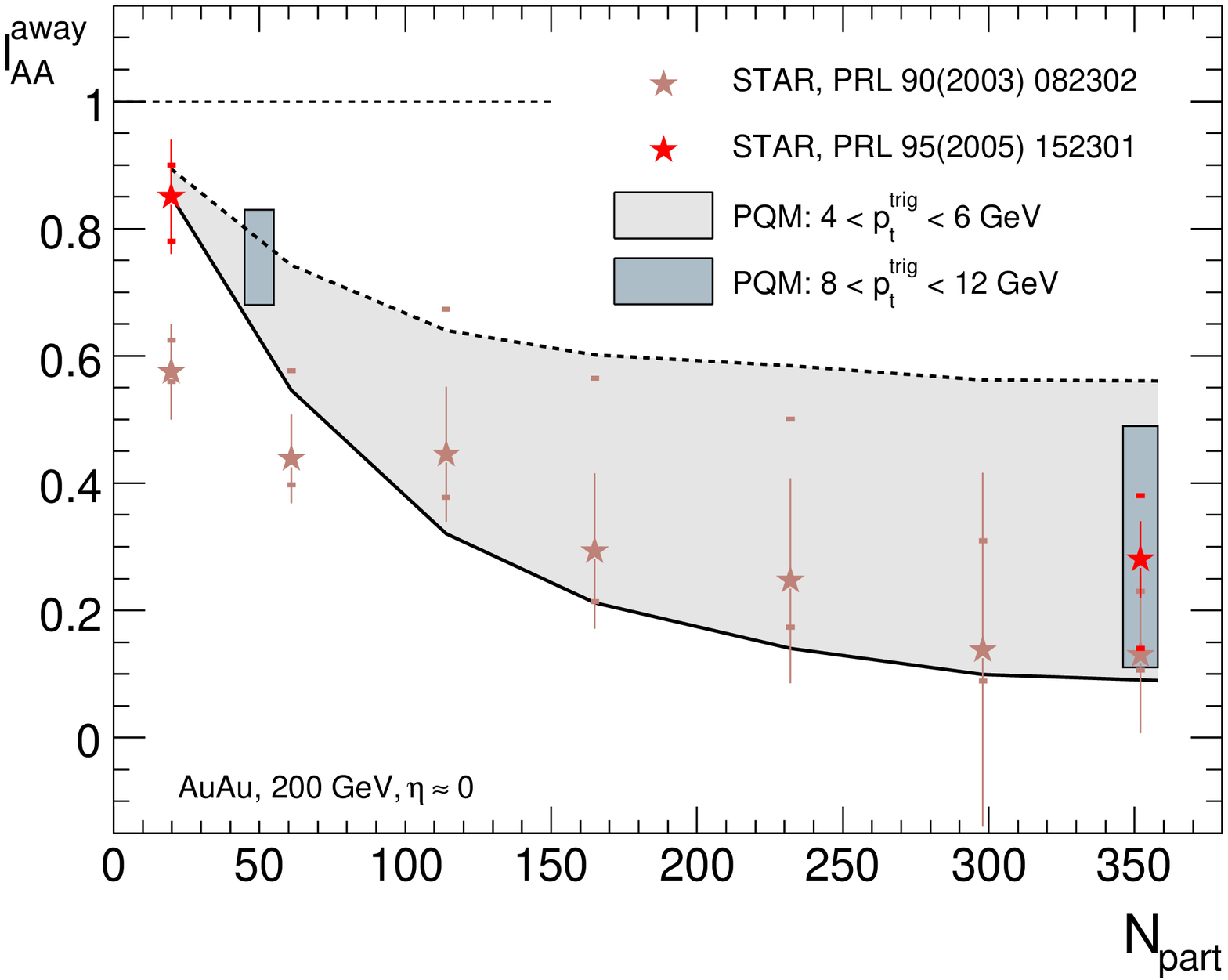}
\vspace{-0.5cm}
\caption[]{$\IAAa(N_{\rm part})$ in \AuAu\ at $\snn=200~\gev$. Data are for 
           $4<\ptt\le6~\gev$ and $2~\gev\le\pta\le\ptt$ with statistical 
           (bars) and systematic (ticks) errors~\cite{cl:starIAA}.
           PQM results for different $\pt$ trigger ranges are shown.
\label{cl:fig2}}
\end{minipage}
\end{figure}

Typically, the leading-particle suppression is quantified via the 
nuclear modification factor,
\begin{equation}      
  \label{cl:eq:raa}
  R_{\rm AA}(\pt,\eta) \equiv 
  \frac{1}{\av{N_{\rm coll}}_{\rm centrality\,class}} \times
  \frac{{\rm d}^2 N_{\rm AA}/{\rm d}\pt{\rm d}\eta}
  {{\rm d}^2 N_{\rm pp}/{\rm d}\pt{\rm d}\eta} \,,  
\end{equation}
as the ratio of the yield in AA over the binary-scaled yield in pp 
for a given centrality class. At mid-rapidity, in 
\AuAu\  collisions at $\snn=200~\gev$, $\RAA$ is found to decrease from 
peripheral ($\RAA\simeq 1$) to central events ($\RAA\simeq 0.2$), for 
$\pt\gsim 5~\gev$~(see \fig{cl:fig1}). 

Figure~\ref{cl:fig1} shows the results of our calculation for $\RAA$.
The single parameter of the model is chosen in order to match the 
suppression measured in 0--10\% central \AuAu\ collisions at 
$\snn=200~\gev$~\cite{cl:starRAA,cl:phenixRAA} leading
to an average transport coefficient 
of about $\av{\hat{q}}=14~\gev^2/\fm$ in order to describe the data
---where the average is done over all produced hard partons.   
For partons with initial $\pt\simeq 10~\gev$, the mean energy loss per unit 
path length is as large as ${\rm d}E/{\rm d}x\simeq 2~\gev/\fm$. 
The probe interacts much stronger with the medium than expected on 
perturbative grounds~\cite{cl:fragile}. This limits the sensitivity 
to $\hat{q}$~\cite{cl:fragile} and leads to 
{\it surface emission}~\cite{cl:PQM,cl:mullerjetpheno}: 
we find that surviving partons yielding hadrons with $\pt>5~\gev$ 
are, on average, emitted from a depth of about $1.7~\fm$ and suffer an
energy loss of less than $0.3~\gev/\fm$.

In light of this observation, it is interesting to consider the suppression 
of back-to-back jet-like two-particle correlations. 
The magnitude of the suppression is usually quantified by the factor 
$\IAAa = D^{\rm away}_{\rm AA}/D^{\rm away}_{\rm pp}$,
where the di-hadron correlation strength, $D^{\rm away}_{\rm pp(AA)}$,
for an associated hadron, $h_2$, with $p_{\rm t,2}$ in the opposite 
azimuthal direction from a trigger hadron, $h_1$, with $p_{\rm t,1}$,
\begin{equation}
\label{cl:eq:iaa}
D^{\rm away}_{\rm pp(AA)} =
\int_{p^{\rm trig}_{\rm t,min}}^{p^{\rm trig}_{\rm t,max}}\d p_{\rm t,1}
\int_{p^{\rm assoc}_{\rm t,min}}^{p^{\rm assoc}_{\rm t,max}}\d p_{\rm t,2}
\int_{\rm{away\, side}}\d\Delta\phi\,
\frac{\d^3\sigma_{\rm pp(AA)}^{h_1h_2}/\d p_{\rm t,1}\d p_{\rm t,2}
\d\Delta\phi}
{\d\sigma_{\rm pp(AA)}^{h_1}/\d p_{\rm t,1}}\,,
\end{equation}
is integrated over the considered trigger- and associated- 
$\pt$ intervals. Similarly to $\RAA$,  
in \AuAu\ collisions at $\snn=200~\gev$, $\IAAa$ is found to
decrease with increasing centrality, down to about $0.2$--$0.3$ 
for the most central events: see STAR data~\cite{cl:starIAA} 
for \mbox{$4<\ptt<6~\gev$} in \fig{cl:fig2}. 
In the figure we present the results of the PQM calculation 
for the parameter value needed to describe $\RAA$. 
The result is consistent with the data, but it has a large 
uncertainty due to the procedure used to treat the cases in which 
the calculated energy loss is of the order of the parton initial energy. 
We show also our prediction for higher transverse momentum of the trigger 
particle, \mbox{$8<\ptt<12~\gev$}. 
It is consistent with preliminary STAR 
data~\cite{cl:magestroqm2005} reporting $\IAAa\simeq 0.25$, 
normalized to \dAu,
for \mbox{$8<\ptt<15~\gev$} in 0--5\% central collisions 
(data not shown in \fig{cl:fig2}).
\begin{figure}[tb]
\mbox{
\ifbandw
\includegraphics[width=4.1cm]{cL1vsL2-2to15trig-2to15ass-nomedium-bw}
\else
\includegraphics[width=4.1cm]{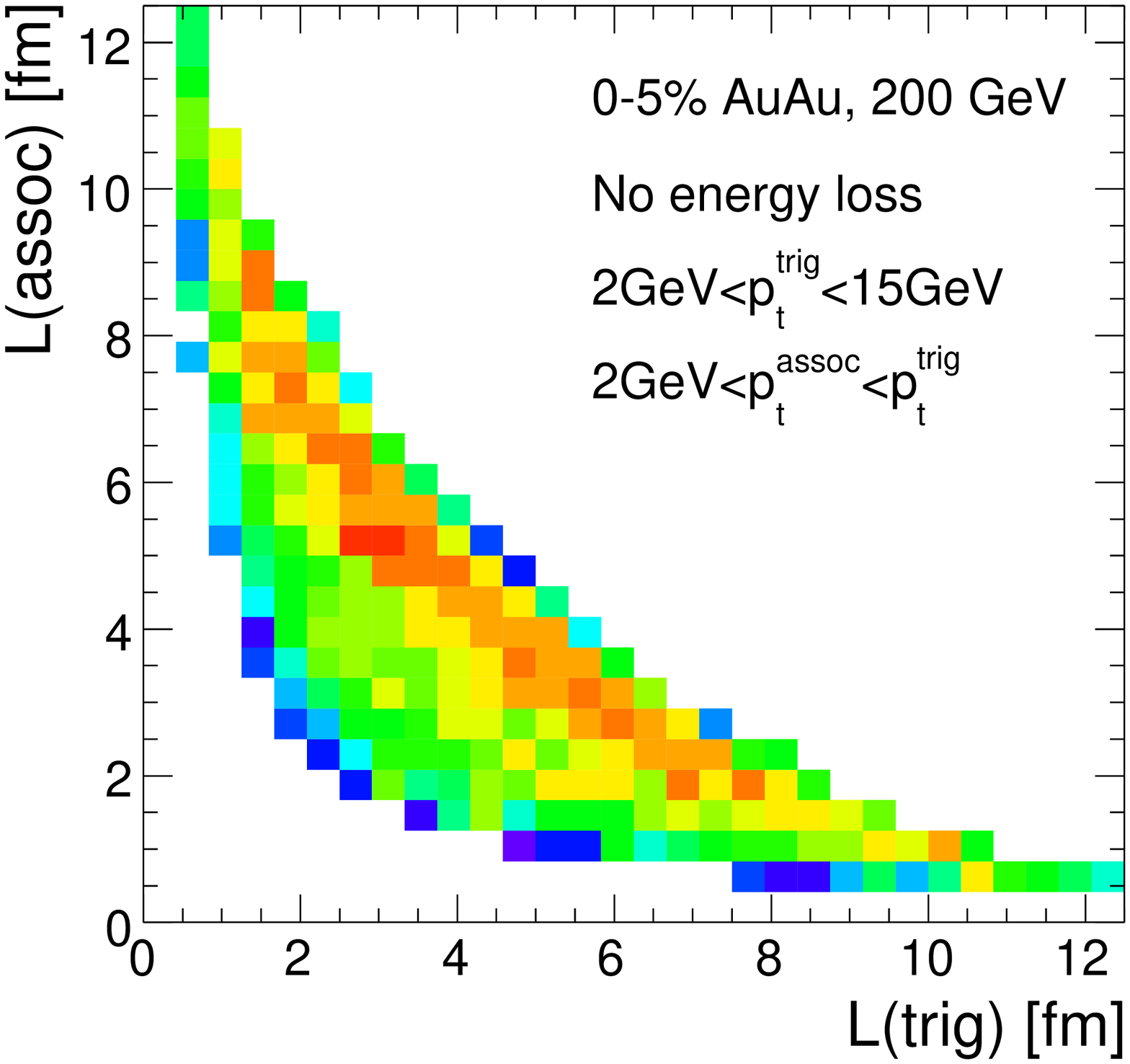}
\fi
\hspace{0.1cm}
\ifbandw
\includegraphics[width=4.1cm]{cL1vsL2-4to6trig-2to6ass-k5e6-nrw-bw}
\else
\includegraphics[width=4.1cm]{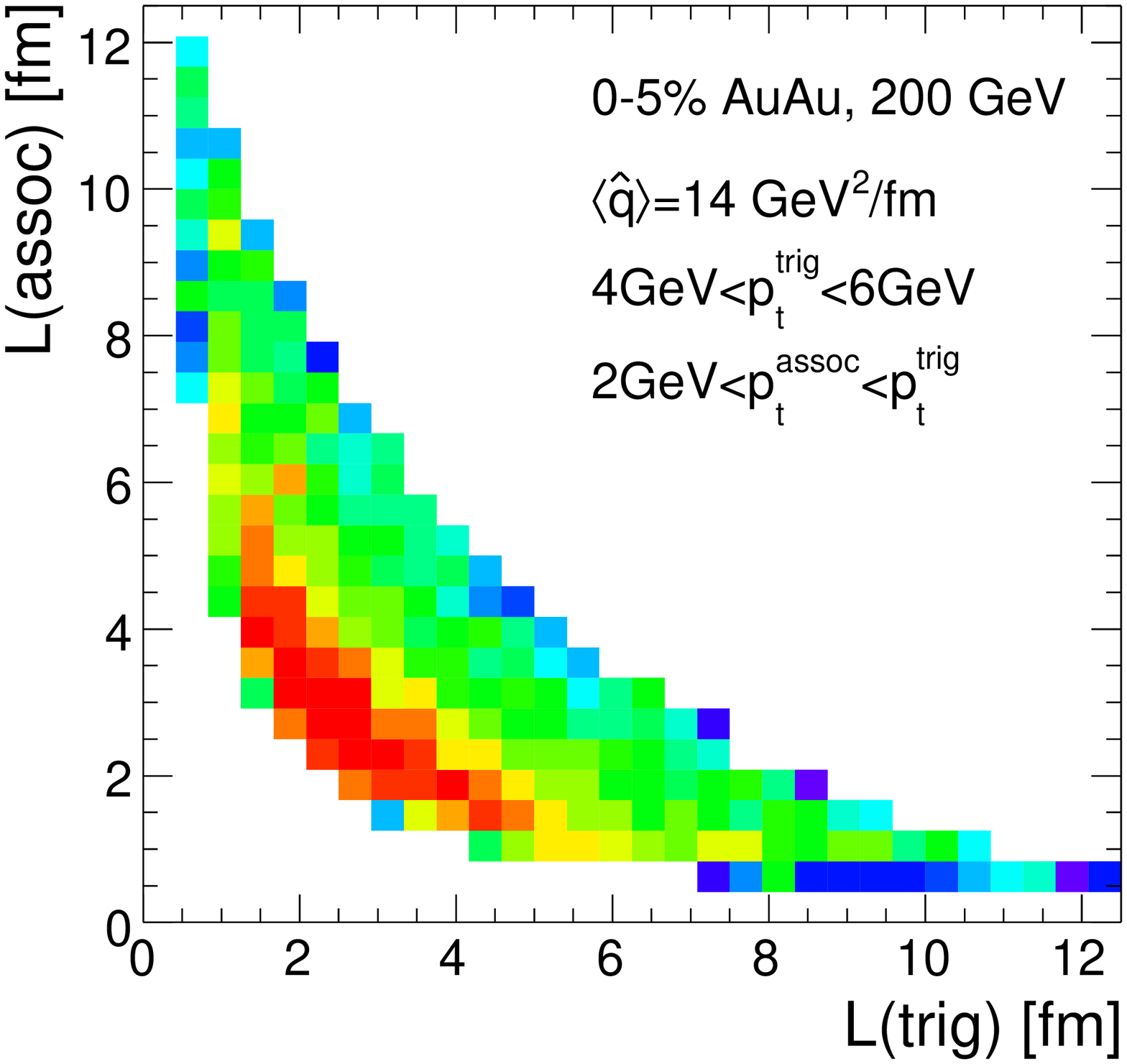}
\fi
\hspace{0.1cm}
\ifbandw
\includegraphics[width=4.1cm]{cL1vsL2-8to15trig-6to15ass-k5e6-nrw-bw}
\else
\includegraphics[width=4.1cm]{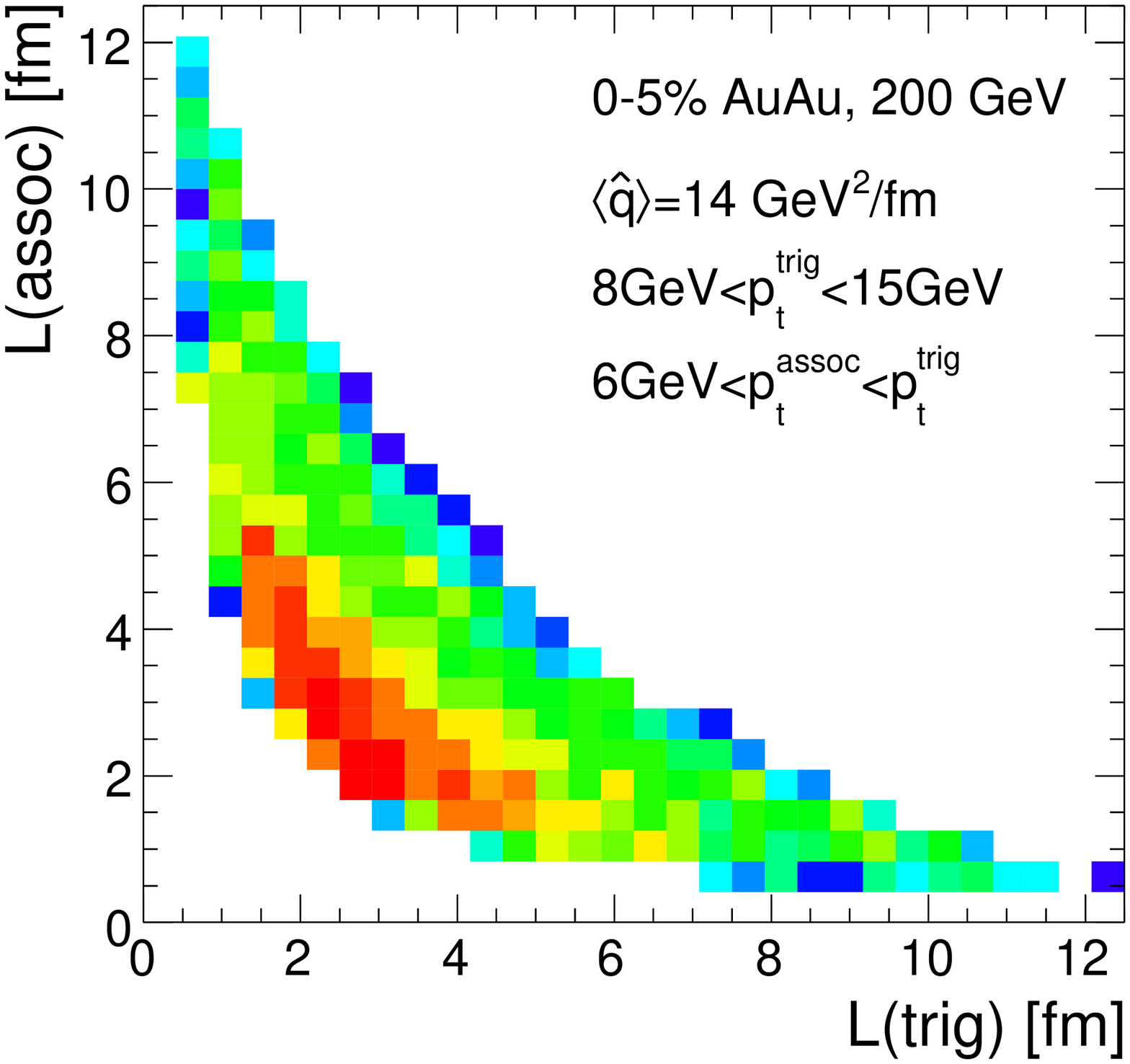}
\fi
}
\vfill
\vspace{0.3cm}
\mbox{
\ifbandw
\includegraphics[width=4.1cm]{cRvsDeltaPhi-2to15trig-2to15ass-nomedium-bw}
\else
\includegraphics[width=4.1cm]{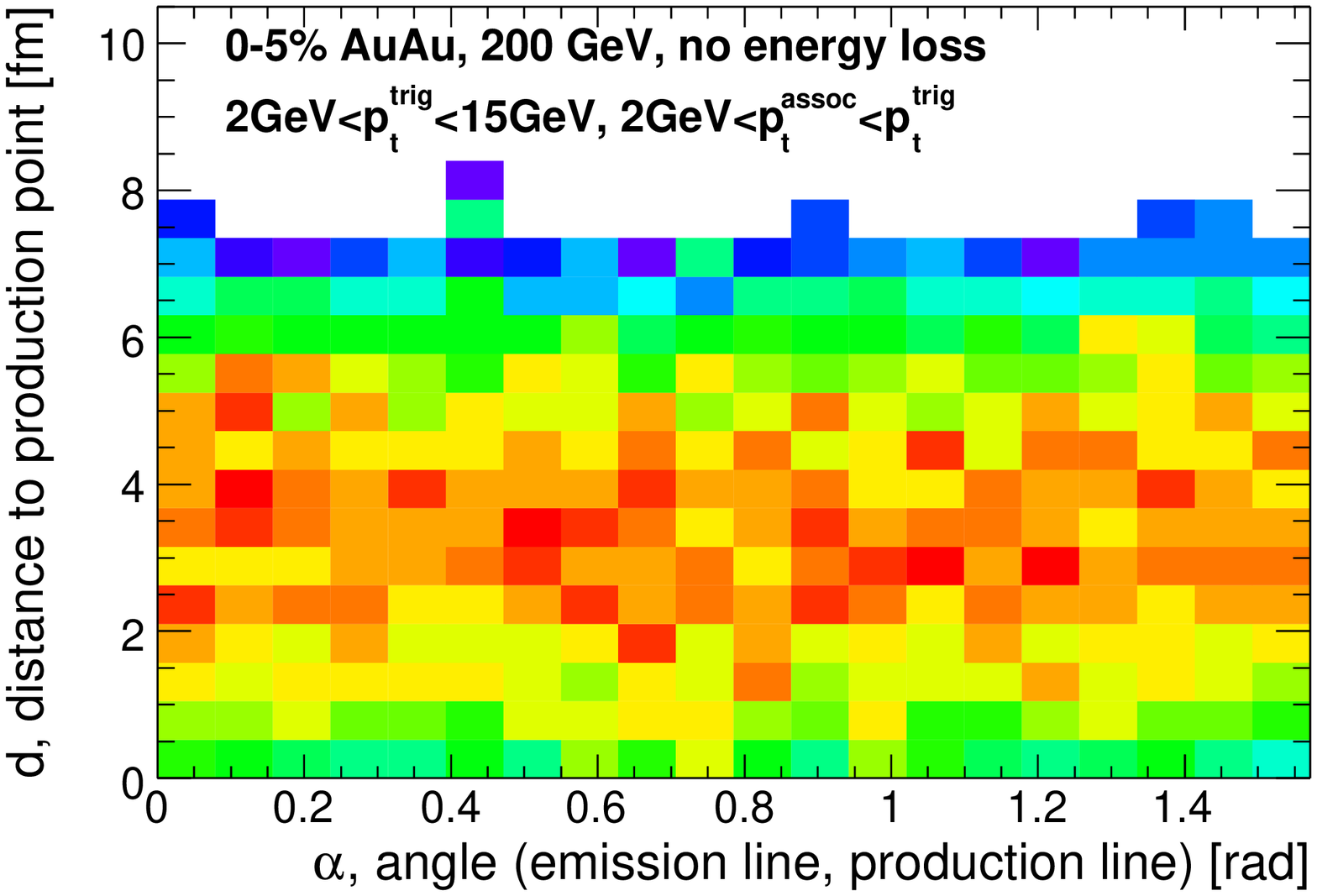}
\fi
\hspace{0.1cm}
\ifbandw
\includegraphics[width=4.1cm]{cRvsDeltaPhi-4to6trig-2to6ass-k5e6-nrw-bw}
\else
\includegraphics[width=4.1cm]{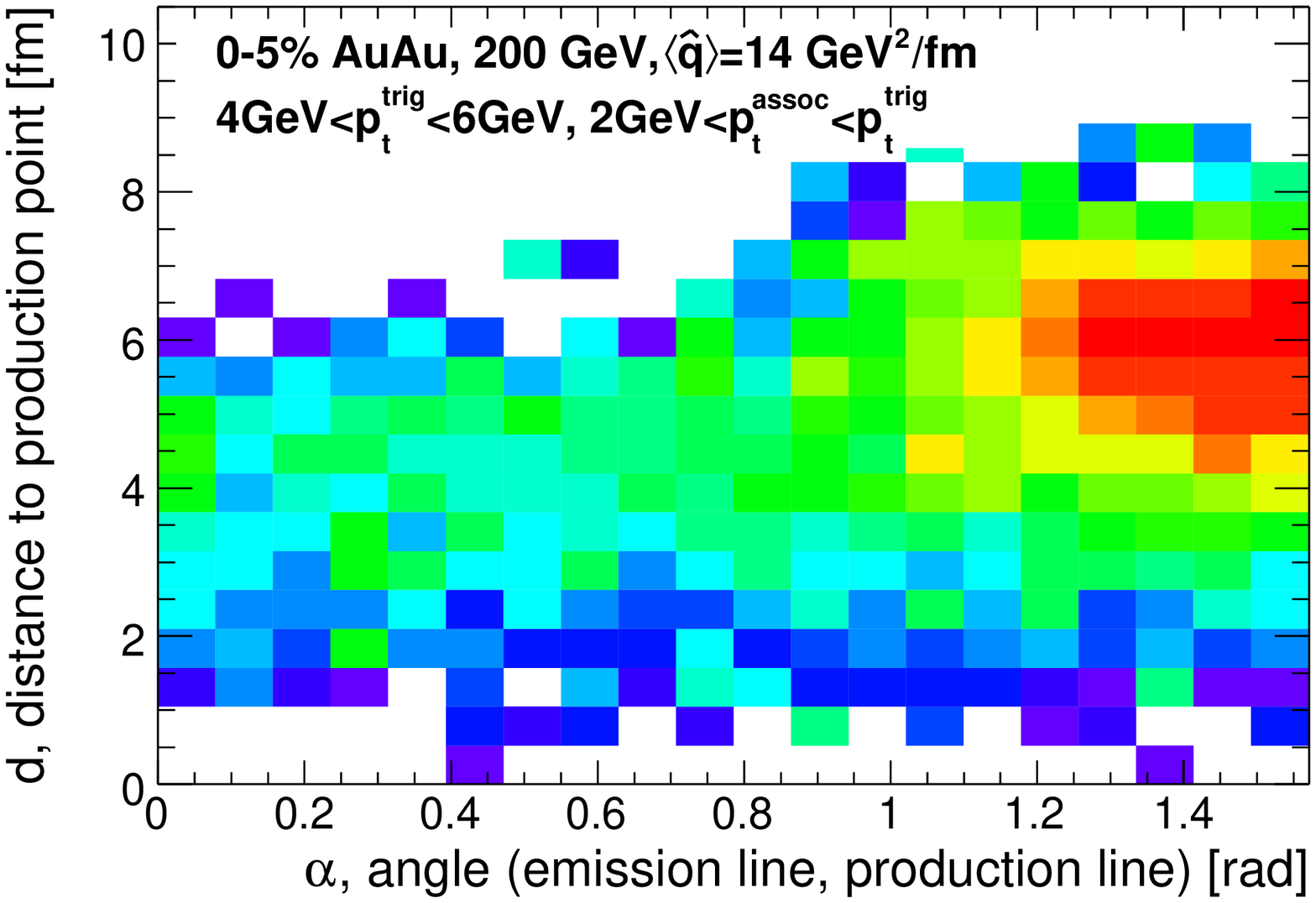}
\fi
\hspace{0.1cm}
\ifbandw
\includegraphics[width=4.1cm]{cRvsDeltaPhi-8to15trig-6to15ass-k5e6-nrw-bw}
\else
\includegraphics[width=4.1cm]{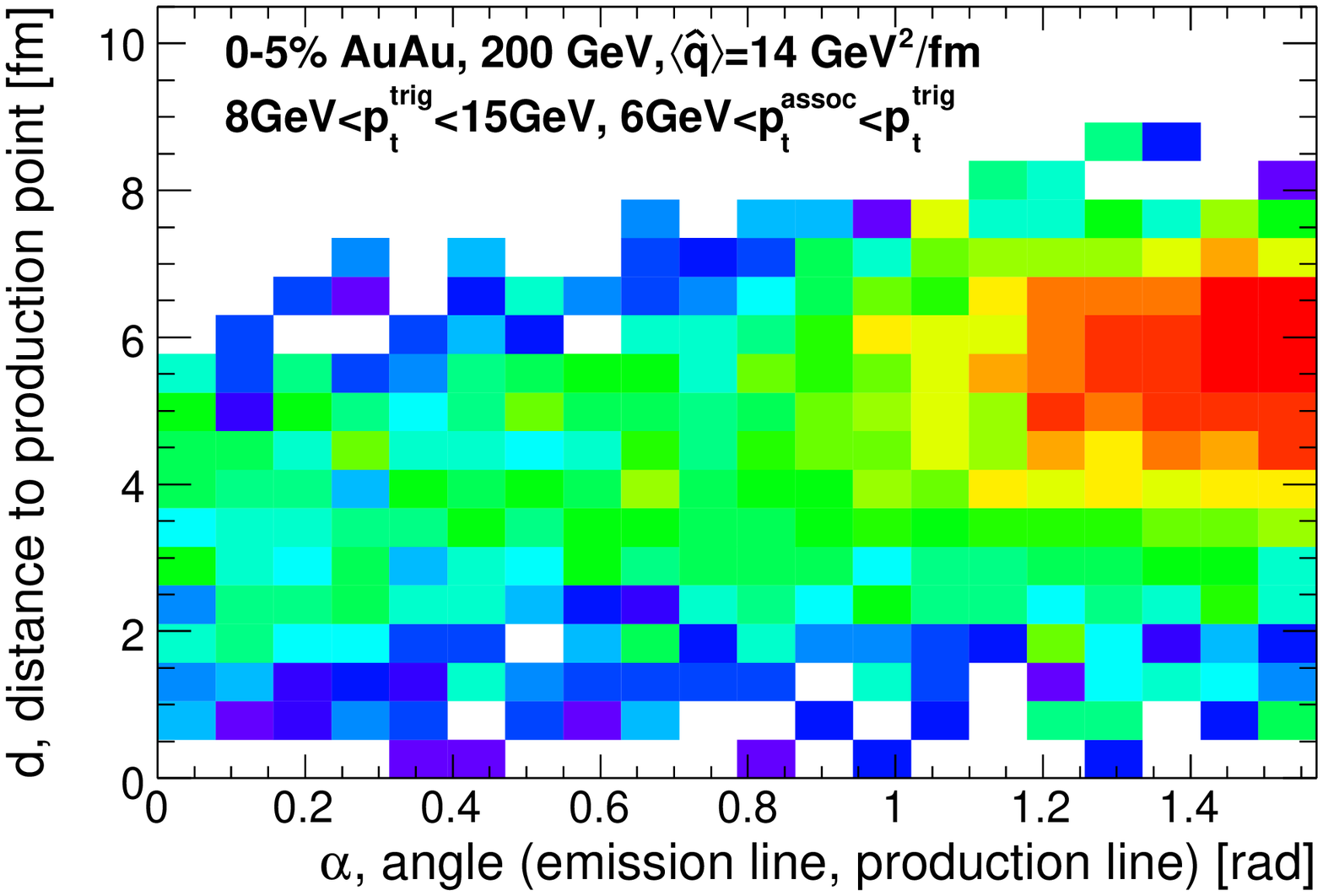}
\fi
}
\vspace{-0.8cm}
\caption[]{(color online)
           Correlation of path-lengths (top) and emission phase-space 
            ---distance from the centre to production point 
             and angle between emission direction and radial direction---
           (bottom) for surviving pairs of back-to-back partons for different 
           conditions (no medium and $\av{\hat{q}}=14~\gev^2/\fm$), yielding 
           hadrons within the reported selection cuts in 0--5\% central 
           \AuAu\ collisions  at $\snn=200~\gev$.
\label{cl:fig3}}
\end{figure}

In \fig{cl:fig3} (top) we show the correlation of the two in-medium 
path lengths for parton pairs yielding hadrons within different 
$\pt$ ranges (see figure).
Without energy loss 
(left panel) the mean parton path length is about $4.5~\fm$,
while it reduces to about $3~\fm$ for $\av{\hat{q}}=14~\gev^2/\fm$,
both for low (central panel) and high (right panel) $\pt$ cuts. 
Note that the `banana' is symmetric: within our model, 
we find no strong difference in the thickness of traversed 
medium between the trigger and the recoil parton. This result 
is illustrated in the bottom panels of \fig{cl:fig3}, where we 
plot the correlation of the distance, $d$, from the centre 
of the overlap region to the pair production point with the 
azimuthal angle, $\alpha$, between the emission direction 
and the radial direction. 
When energy loss is included, the back-to-back parton pairs that emerge 
are those produced {\it close to the medium surface} ($d\simeq 4$--$6~\fm$)
and with propagation direction {\it oriented tangentially} 
with respect to the medium ($\alpha\simeq \pi/2$).

For both sets of trigger cuts, the mean energy loss suffered 
by the surviving away-side partons is less than $0.3~\gev/\fm$,
which is similar to the the single inclusive case, and 
in agreement with the experimental observation of unmodified 
(relative to \dAu) hadron-triggered fragmentation functions~\cite{cl:magestroqm2005}.
In addition, we find no qualitative difference between the two sets 
of trigger cuts (central and right panels of~\fig{cl:fig3}.
This is compatible with the $\IAAa$ values measured by STAR,
which are similar in the two cases~\cite{cl:starIAA,cl:magestroqm2005}.

\section{Conclusions}
\label{concl}
We have discussed jet quenching effects at the top RHIC energy within
the Parton Quenching Model, which combines energy loss calculations 
\`a la BDMPS and a Glauber-based implementation of the collision geometry.
After tuning the single free parameter, the model describes 
(i) the $\pt$-independence of $\RAA$ at high $\pt$, (ii) the centrality 
dependence of $\RAA$, and (iii) the magnitude and centrality dependence of 
the away-side suppression factor $\IAAa$.
Our analysis suggests that the production of high-$\pt$ hadrons in 
central nucleus--nucleus collisions is surface-dominated, not only 
for single hadrons, but also for back-to-back di-jets.

~\\
\noindent
Fruitful discussion with P.~Jacobs, C.A.~Salgado and U.A.~Wiedemann 
are acknowledged.

\vfill\eject
\end{document}